\documentclass[aps,pra,reprint,superscriptaddress,longbibliography]{revtex4-1}
\usepackage{graphicx}
\usepackage{siunitx}
\usepackage[utf8]{inputenc}
\usepackage{newunicodechar}
\usepackage{float}
\usepackage{amsmath}

\newunicodechar{χ}{$\chi^{(3)}$}

\begin{document}

\title{Quasi-phase-matched supercontinuum-generation in photonic waveguides}

\newcommand{\NISTTF}{Time and Frequency Division, National Institute of Standards and Technology, Boulder, Colorado 80305, U.S.A.}
\newcommand{\NISTG} {Center for Nanoscale Science and Technology, NIST, Gaithersburg, Maryland 20899, U.S.A.}
\newcommand{\Gustavus} {Gustavus Adolphus College, Saint Peter, Minnesota 56082, U.S.A.}
\newcommand{\UCSB}  {Department of Electrical and Computer Engineering, University of California, Santa Barbara, California 93106, U.S.A.}
\newcommand{\CU}    {Department of Physics, University of Colorado, Boulder, Colorado, 80309, U.S.A.}

\author{Daniel~D.~Hickstein} \email[]{danhickstein@gmail.com} \affiliation{\NISTTF}
\author{Grace~C.~Kerber}     \affiliation{\NISTTF} \affiliation{\Gustavus}
\author{David~R.~Carlson}    \affiliation{\NISTTF}
\author{Lin~Chang}           \affiliation{\UCSB}
\author{Daron~Westly}        \affiliation{\NISTG} 
\author{Kartik~Srinivasan}   \affiliation{\NISTG} 
\author{Abijith~Kowligy}     \affiliation{\NISTTF}
\author{John~E.~Bowers}      \affiliation{\UCSB}
\author{Scott~A.~Diddams}    \affiliation{\NISTTF} \affiliation{\CU} 
\author{Scott~B.~Papp}       \affiliation{\NISTTF} \affiliation{\CU}

\date{\today}

\begin{abstract}
Supercontinuum generation in integrated photonic waveguides is a versatile source of broadband light, and the generated spectrum is largely determined by the phase-matching conditions. Here we show that quasi-phase-matching via periodic modulations of the waveguide structure provides a useful mechanism to control the evolution of ultrafast pulses during supercontinuum generation. We experimentally demonstrate quasi-phase-matched supercontinuum to the TE\textsubscript{20} and TE\textsubscript{00} waveguide modes, which enhances the intensity of the SCG in specific spectral regions by as much as 20~dB. We utilize higher-order quasi-phase-matching (up to the 16\textsuperscript{th} order) to enhance the intensity in numerous locations across the spectrum. Quasi-phase-matching adds a unique dimension to the design-space for SCG waveguides, allowing the spectrum to be engineered for specific applications.
\end{abstract}

\maketitle


Supercontinuum generation (SCG) is a $\chi^{(3)}$ nonlinear process where laser pulses of relatively narrow bandwidth can be converted into a continuum with large spectral span \cite{alfano_supercontinuum_2016, dudley_supercontinuum_2006, agrawal_nonlinear_2007}. SCG has numerous applications, including self-referencing frequency combs \cite{jones_carrier-envelope_2000, holzwarth_optical_2000,diddams_direct_2000}, microscopy \cite{betz_excitation_2005}, spectroscopy \cite{coddington_dual-comb_2016}, and tomography \cite{moon_ultra-high-speed_2006}. SCG is traditionally accomplished using bulk crystals or nonlinear fiber, but recently, ``photonic waveguides'' (on-chip waveguides produced using nanofabrication techniques) have proven themselves as a versatile platform for SCG, offering small size, high nonlinearity, and increased control over the generated spectrum \cite{klenner_gigahertz_2016, porcel_two-octave_2017, mayer_frequency_2015, epping_chip_2015, kuyken_octave-spanning_2015-1, boggio_dispersion_2014, carlson_photonic-chip_2017, carlson_self-referenced_2017, oh_coherent_2017, hickstein_ultrabroadband_2017}. The spectral shape and efficiency of SCG is determined by the input pulse parameters, the nonlinearity of the material, and the refractive index of the waveguide, which determines the phase-matching conditions.

\begin{figure}[b!]
	\includegraphics[width=\linewidth]{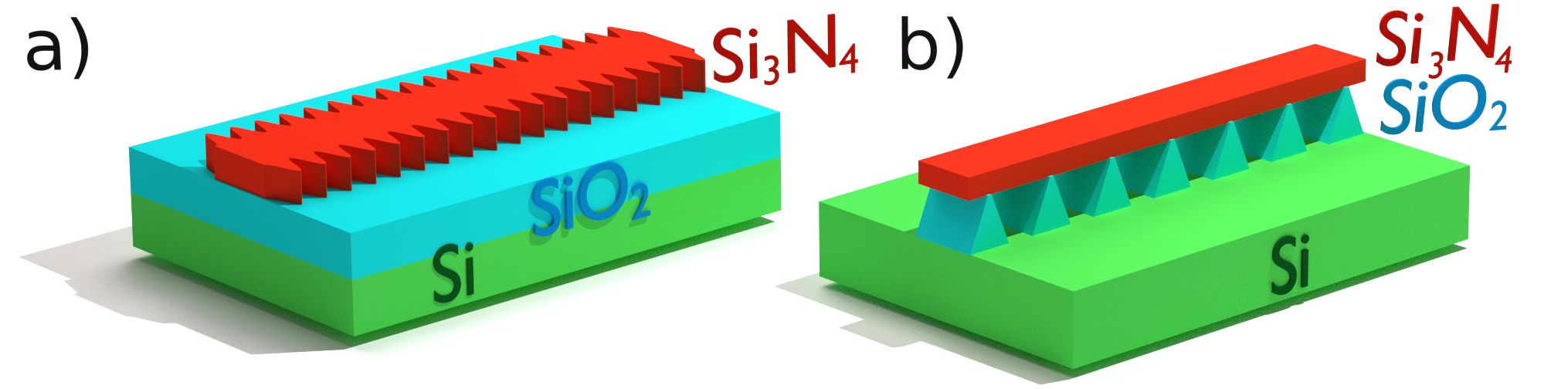}
    \includegraphics[width=\linewidth]{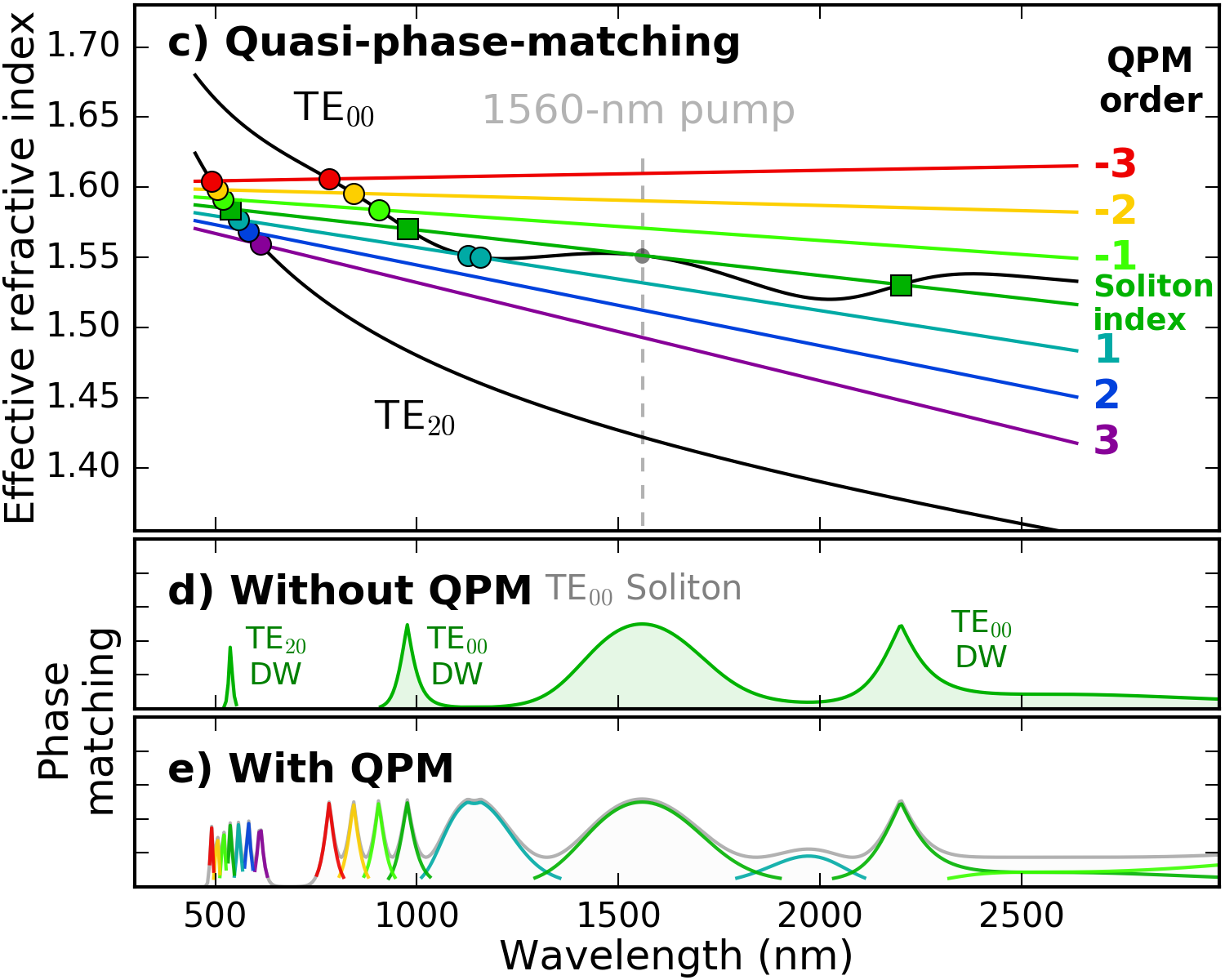}
	\caption{\label{fig:overview} a,b) Quasi-phase-matching (QPM) of supercontinuum generation in on-chip photonic waveguides can be achieved via waveguide-width modulation (a) or cladding-modulation (b). c) When the effective index of the soliton in the $\mathrm{TE_{00}}$ mode (``soliton index'') intersects the effective index of a waveguide mode (black curves), phase matching to dispersive waves (DWs) is achieved (green squares) and a spectrum with several peaks (d) is generated. The periodic modulation of the waveguide can enable numerous QPM orders, both positive and negative, which can allow QPM-DW generation to the fundamental mode and to higher-order modes (circles), producing a spectrum with many peaks~(e). Note: the index curvature is exaggerated to better show phase-matching.}
\end{figure}

Specifically, when phase-matching between a soliton and quasi-continuous-wave (CW) light is achieved, strong enhancements of the intensity of the supercontinuum spectrum can occur in certain spectral regions. These spectral peaks are often referred to as a dispersive waves (DWs) \cite{dudley_supercontinuum_2006, driben_resonant_2015, mclenaghan_few-cycle_2014, akhmediev_cherenkov_1995}, and they are often crucial for providing sufficient spectral brightness for many applications. The soliton-DW phase-matching condition is typically satisfied by selecting a material with a favorable refractive index profile and engineering the dimensions of the waveguide to provide DWs at the desired wavelengths \cite{boggio_dispersion_2014, carlson_photonic-chip_2017}. However, there are limitations to the refractive index profile that can be achieved by adjusting only the waveguide cross-section. Quasi-phase-matching (QPM) takes a different approach, utilizing periodic modulations of the material nonlinearity to achieve an end-result similar to true phase-matching \cite{boyd_nonlinear_2008, franken_optical_1963, armstrong_interactions_1962, fejer_quasi-phase-matched_1992}. QPM is routinely employed to achieve high conversion efficiency for nonlinear processes such as second harmonic generation and different frequency generation, and QPM can also be used to satisfy the phase-matching conditions for DW generation \cite{kudlinski_parametric_2015, luo_resonant_2015, conforti_multiple_2016, wright_ultrabroadband_2015, droques_dynamics_2013}.

Here we show that periodic modulations of the effective mode area can enable QPM of DW generation in photonic silicon nitride ($\mathrm{Si_3N_4}$) waveguides, enhancing the intensity of the supercontinuum in specific spectral regions determined by the modulation period. Experimentally, we utilize a sinusoidal modulation of the waveguide width to enable first-order QPM to the $\mathrm{TE_{20}}$ mode. Additionally, we demonstrate that periodic $\mathrm{SiO_2}$ under-cladding provides numerous orders of QPM to both the $\mathrm{TE_{20}}$ and $\mathrm{TE_{00}}$ modes. This quasi-phase-matched dispersive-wave (QPM-DW) scheme provides a fundamentally different approach to phase-matching in SCG, allowing light to be generated outside the normal wavelength range, and providing separate control over the group-velocity dispersion (GVD) of the waveguide (which influences soliton propagation) and DW phase-matching, capabilities that are desirable for many applications of SCG.

In the regime of anomalous GVD, the nonlinearity of the material can balance GVD and allow pulses to propagate while remaining temporally short. Such solitons can propagate indefinitely, unless perturbed \cite{agrawal_nonlinear_2007, dudley_supercontinuum_2006}. However, in the presence of higher-order dispersion, some wavelengths of quasi-CW light may propagate at the same phase velocity as the soliton. Light at these wavelengths can ``leak out'' of the soliton, in the form of DW radiation \cite{dudley_supercontinuum_2006}, which is also referred to as ``resonant radiation'' \cite{driben_resonant_2015, mclenaghan_few-cycle_2014} or ``optical Cherenkov radiation'' \cite{akhmediev_cherenkov_1995}. In the SCG process, higher-order solitons undergo soliton fission and can convert significant amounts of energy into DW radiation \cite{dudley_supercontinuum_2006}. The phase-matching condition for DW generation (in the absence of QPM) is simply \cite{akhmediev_cherenkov_1995}
\begin{equation}
\label{eq:dw}
n(\lambda_\mathrm{s}) + (\lambda - \lambda_\mathrm{s}) \frac{dn}{d\lambda}(\lambda_\mathrm{s}) + \gamma \, p \, \lambda = n(\lambda),
\end{equation}
where $\lambda$ is wavelength, $\lambda_\mathrm{s}$ is the center wavelength of the soliton, $n$ is the effective index of the waveguide, $\frac{dn}{d\lambda}(\lambda_\mathrm{s})$ is the slope of the $n$-versus-$\lambda$ curve evaluated at $\lambda_\mathrm{s}$, $\gamma$ is the effective nonlinearity of the waveguide, and $p$ is the peak power. The left side of Eq.~\ref{eq:dw} represents the effective index of the soliton while the right side represents the effective index of the DW. This equation has a simple graphical interpretation; because all wavelengths in the soliton travel with the same group velocity, the effective index of the soliton is simply a straight line (Fig.~\ref{fig:overview}c). Where this line crosses the refractive index curve for any waveguide mode (black lines in Fig.~\ref{fig:overview}c), DW generation is phase matched.

Periodic modulations of the waveguide change the effective area of the mode, modulating both $\gamma$ and the GVD, enabling QPM-DW generation (see Supplemental Materials, section IV). The contribution of a modulation (with period $\Lambda$) to the wavevector phase-mismatch will be $k_\mathrm{QPM} = 2\pi q/\Lambda$, where $q$ is the QPM order and can be any positive or negative integer. Thus, the phase-matching condition for dispersive wave generation, in the presence of QPM is
\begin{equation}
\label{eq:qpm}
\smash{\underbrace{n(\lambda_\mathrm{s}) + (\lambda-\lambda_\mathrm{s}) \frac{dn}{d\lambda}  (\lambda_\mathrm{s}) + \gamma \, p\lambda_\mathrm{s}}_{\text{Soliton index}}  = \underbrace{n(\lambda)}_{\text{DW index}} + \underbrace{\frac{2\pi q \lambda}{\Lambda}.}_{\text{QPM}}}
\end{equation}
\\
This phase-matching condition has a similar graphical interpretation; an additional line is drawn for each QPM order, with curve-crossings indicating QPM of DWs (Fig.~\ref{fig:overview}c).

In addition to satisfying the QPM condition (Eq.~\ref{eq:qpm}), three additional requirements must be met for efficient DW generation. (1) The QPM order $q$ must be a strong Fourier component of the periodic modulation. (2) There must be overlap between the spatial mode of the soliton and the mode of the DW. (3) The DW must be located in a spectral region where the soliton has significant intensity, a requirement that applies regardless of the phase-matching method.


\begin{figure}
	\includegraphics[width=\linewidth]{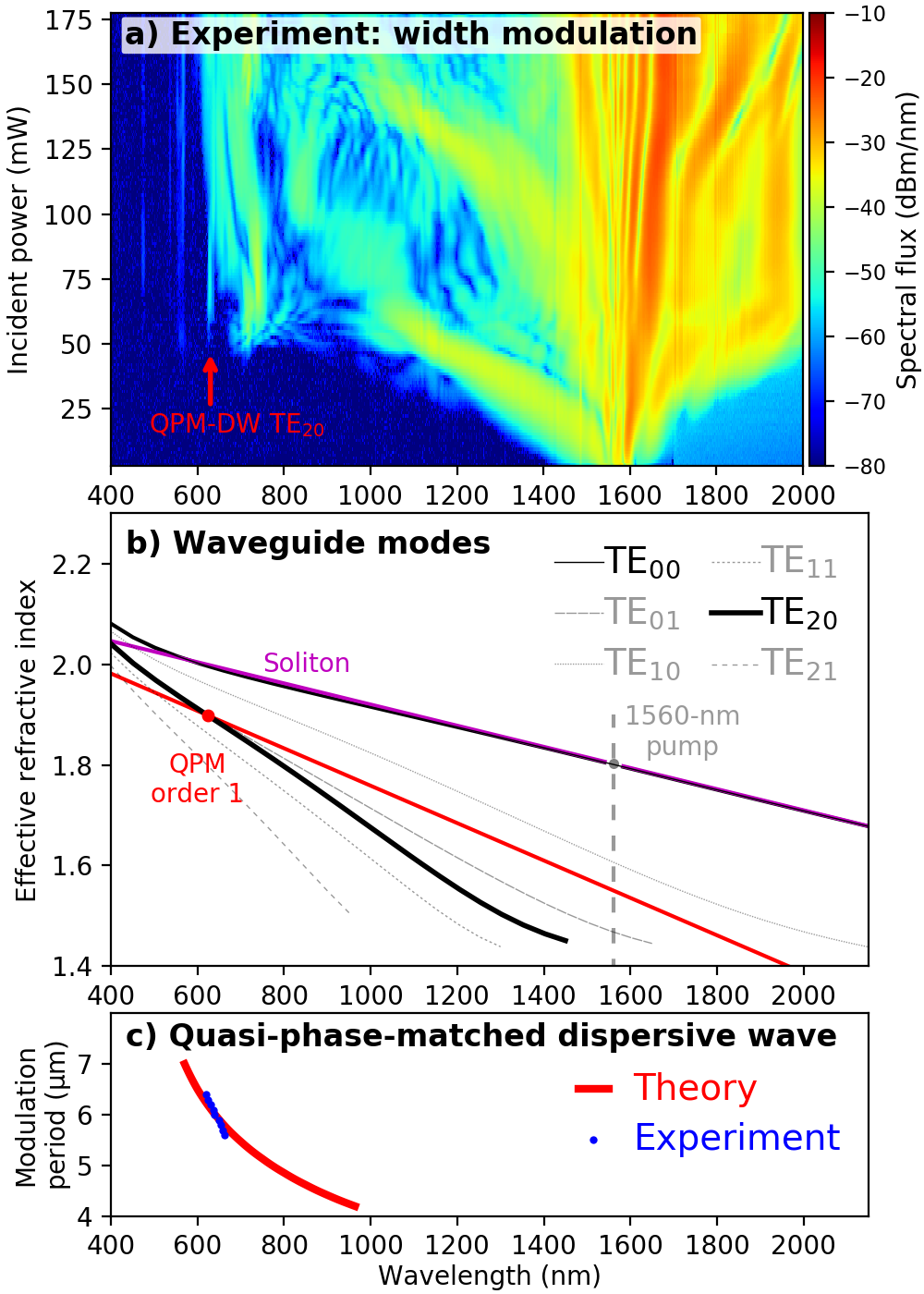}
	\caption{\label{fig:wiggle}
    a)~The spectrum of supercontinuum generation from a width-modulated waveguide as a function of input power. The arrow indicates the quasi-phase-matched dispersive wave (QPM-DW) to the $\mathrm{TE_{20}}$ mode. b)~The effective refractive index of various modes of the waveguide as a function of wavelength. When the index of the soliton including a first order grating effect from the 6.2-\SI{}{\micro\meter} width-modulation (red line) crosses the $\mathrm{TE_{20}}$ mode, a QPM-DW is generated. c)~The calculated spectral location of the $\mathrm{TE_{20}}$ QPM-DW as a function of the width-modulation period is in agreement with experimental results. The slight difference in slope may arise from irregularities in the dimensions of the waveguides.}
\end{figure}

Experimentally, we explore two different approaches for QPM-DW generation in $\mathrm{Si_3N_4}$ waveguides: width-modulated waveguides and cladding-modulated waveguides. The width-modulated $\mathrm{Si_3N_4}$ waveguides (Fig.~\ref{fig:overview}a) are fully $\mathrm{SiO_2}$-clad and have a thickness of 750~nm, a maximum width of 1500~nm, and an overall length of 15~mm. Over a 6-mm central region, the width is modulated sinusoidally from 1250 to 1500~nm. Multiple waveguides are fabricated on the same silicon chip, and each waveguide has a different width modulation period, which ranges from 5.5 to \SI{6.5}{\micro\meter}. Each cladding-modulated waveguide (Fig.~\ref{fig:overview}b) consists of a 700-nm-thick Si$_3$N$_4$ waveguide that is completely air-clad, except for underlying SiO$_2$ support structures. The support structures are placed every \SI{200}{\micro\meter} along the waveguide, and each one contacts the Si$_3$N$_4$ waveguide for approximately \SI{20}{\micro\meter}. For the cladding-modulated waveguides, the modulation period is kept constant, but several waveguide widths are tested, ranging from 3000 to 4000~nm. 

We generate supercontinuum by coupling $\sim$80~fs pulses of 1560-nm light from a compact 100~MHz Er-fiber frequency comb \cite{sinclair_compact_2015}. The power is adjusted using a computer-controlled rotation-mount containing a half-waveplate, which is placed before a polarizer. The polarization is set to horizontal (i.e., parallel to the Si-wafer surface and along the long dimension of the rectangular Si$_3$N$_4$ waveguide), which excites the lowest order quasi-transverse-electric ($\mathrm{TE_{00}}$) mode of the waveguide. We record the spectrum at many increments of the input power using an automated system \cite{hickstein_gracefulosa_2017} that interfaces with both the rotation mount and the optical spectrum analyzers. The waveguide modes (and their effective indices) are calculated using a vector finite-difference modesolver \cite{fallahkhair_vector_2008, bolla_empy_2017}, using published refractive indices for $\mathrm{Si_3N_4}$ \cite{luke_broadband_2015} and $\mathrm{SiO_2}$ \cite{malitson_interspecimen_1965}. Further experimental details are found in the SM.

For the width-modulated waveguides, a narrow peak appears in the spectrum in the 630-nm region (Fig.~\ref{fig:wiggle}a), and the location of this peak changes with the width-modulation period (Fig.~\ref{fig:wiggle}c). By calculating the refractive index of the higher order modes of the waveguide, and including the QPM effect from the periodic width modulation (Fig.~\ref{fig:wiggle}b and Eq.~\ref{eq:qpm}), we find the QPM-DW generation to the $\mathrm{TE_{20}}$ mode is a likely mechanism for the appearance of this peak (Fig.~\ref{fig:wiggle}c). The preference for QPM-DW generation to the $\mathrm{TE_{20}}$ mode is a result of the modal overlap \cite{agrawal_nonlinear_2007, lin_nonlinear_2007} (Fig.~\ref{fig:modes}) between the $\mathrm{TE_{20}}$ mode at the DW wavelength ($\sim$630 nm) and the $\mathrm{TE_{00}}$ mode at the soliton wavelength (1560~nm). In general, modes that are symmetric in both the vertical and horizontal (such as the $\mathrm{TE_{20}}$ mode) will have much higher overlap to the fundamental mode than antisymmetric modes (see SM, section III), and are consequently the most commonly used for model phase-matching schemes \cite{guo_second-harmonic_2016}.

\begin{figure}
    \includegraphics[width=\linewidth]{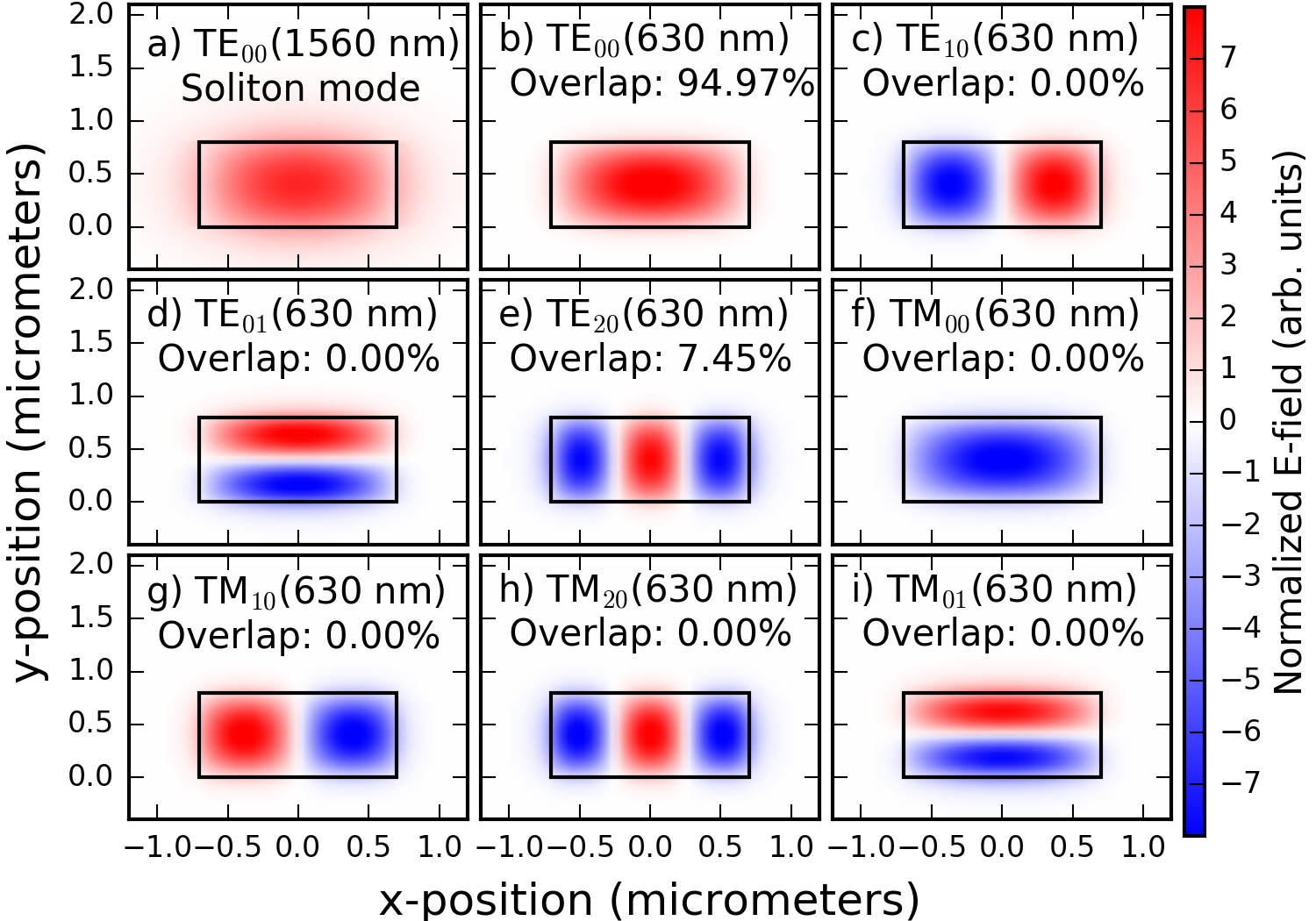}
    \caption{\label{fig:modes}
    Electric field profiles for the waveguide modes for a fully $\mathrm{SiO_2}$-clad $\mathrm{Si_3N_4}$ waveguide. a) The $\mathrm{TE_{00}}$ mode at 1560~nm, which is the expected mode of the soliton. b-i) The electric field for various higher order modes at 630~nm, which is the approximate wavelength for the QPM-DW observed for the width-modulated waveguides. The result of the overlap integral of each mode with the $\mathrm{TE_{00}}$ mode at 1560~nm is listed. Only the $\mathrm{TE_{00}}$ and $\mathrm{TE_{20}}$ modes have overlap integrals that are not vanishingly small. Note: for TM modes, $E_y$ is shown, while $E_x$ is shown for TE modes.}
\end{figure}

\begin{figure}
    \includegraphics[width=\linewidth]{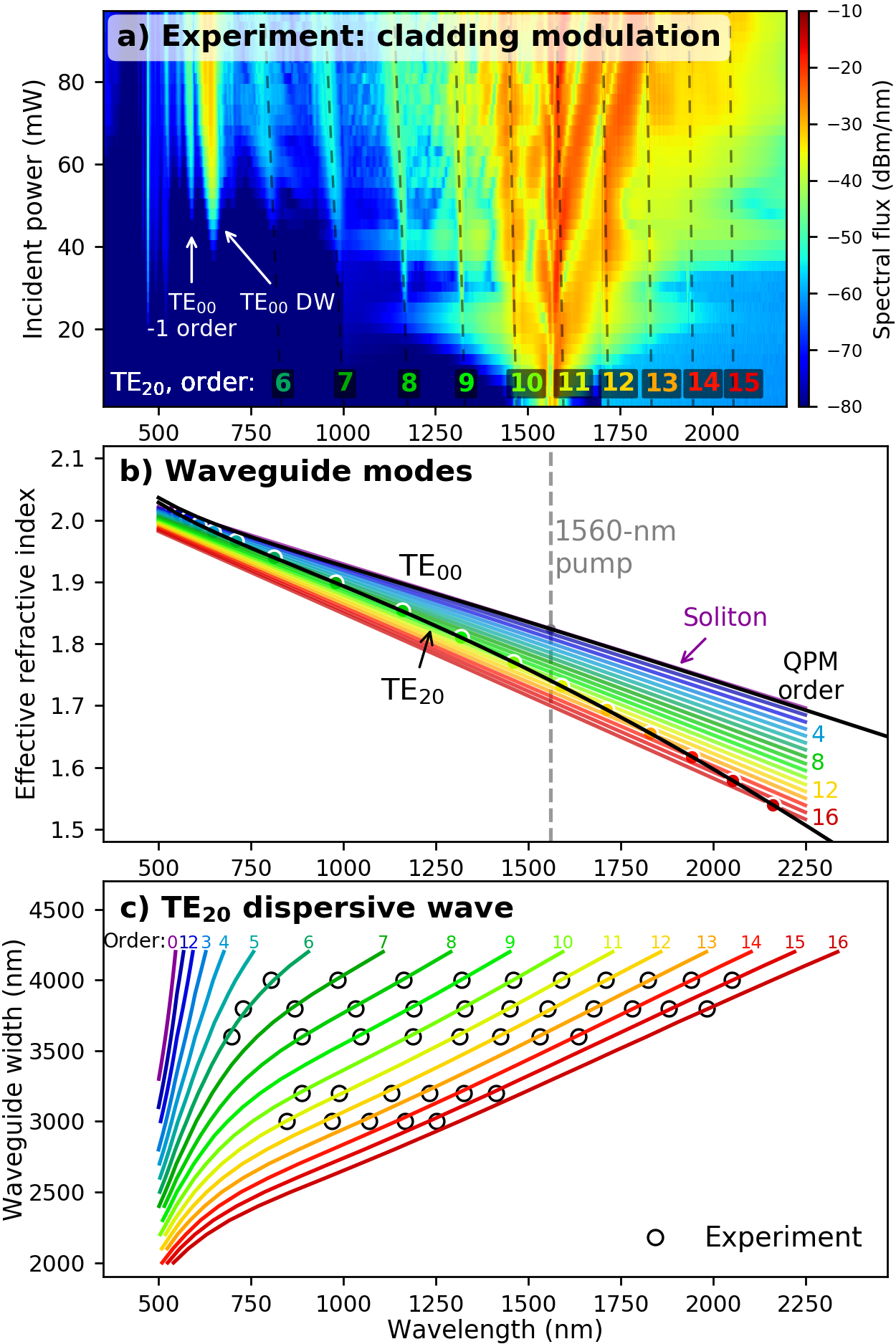}
    \caption{\label{fig:suspended} a) Supercontinuum generation from a 4000-nm-width cladding-modulated waveguide, showing many QPM-DW peaks resulting from QPM orders 6-to-15 to the $\mathrm{TE_{20}}$ mode. The locations of the QPM-DWs predicted by theory (dashed lines) agree with the experiment. DWs corresponding to phase-matching to the soliton and to --1 order QPM to the $\mathrm{TE_{00}}$ mode are indicated with arrows. b) The effective index of the $\mathrm{TE_{00}}$ and $\mathrm{TE_{20}}$ modes, compared with the effective index of the soliton for various QPM orders (colorful lines). The dots indicate the location of the QPM-DW in the $\mathrm{TE_{20}}$ mode. c) Calculations indicate that the locations of the QPM-DWs change as a function of waveguide width, in agreement with experiment. Note: the sharp peaks in the 530-nm region of (a) are a result of third-harmonic generation to higher-order spatial modes \cite{carlson_self-referenced_2017}.}
\end{figure} 

For the cladding-modulated waveguides, many QPM-DW peaks are seen in the supercontinuum spectrum (Fig~\ref{fig:suspended}a). In some cases, the enhancement in the spectral intensity is as high as 20~dB. Similarly to the width-modulated waveguides, an analysis of the refractive index profile indicates that the $\mathrm{TE_{20}}$ mode is responsible for the QPM-DW generation (Fig.~\ref{fig:suspended}). Interestingly, peaks are observed corresponding to both odd- and even-order QPM, and effects up to the 16$^\mathrm{th}$ QPM order are detectable. This situation differs from the preference for low, odd-ordered QPM effects in typical QPM materials (such as periodically poled lithium niobate, PPLN \cite{fejer_quasi-phase-matched_1992}), which usually employ a 50-percent duty-cycle modulation. In contrast, the cladding-modulated $\mathrm{Si_3N_4}$ waveguides have short regions of oxide cladding, followed by long regions of fully air-clad $\mathrm{Si_3N_4}$. This high-duty-cycle square wave is composed of both even- and odd-order harmonics, and consequently provides both even- and odd-order QPM. The simulated QPM-DW positions are in good agreement with the experiment for all grating orders and waveguide widths. Importantly, we see that QPM can still produce strong DWs with QPM orders of 8 or more, indicating that QPM for strongly phase-mismatched processes could be achieved with higher-order QPM instead of short modulation periods, potentially avoiding fabrication difficulties and scattering loss. We also observe QPM-DW generation to the $\mathrm{TE_{00}}$ mode (Fig.~\ref{fig:suspended}a), which is reproduced by numerical simulations using the nonlinear schr\"odinger equation \cite{heidt_efficient_2009,hult_fourth-order_2007,ycas_pynlo_2016} (See Supplementary Material, Fig.~S2).


This is the first demonstration of QPM to produce DWs in on-chip waveguides, but it is interesting to note that the QPM-DWs have been observed in a variety of situations. Indeed, the ``Kelly sidebands'' \cite{kelly_characteristic_1992} seen in laser cavities and sidebands seen during soliton propagation in long-distance fiber links \cite{matera_sideband_1993} are both examples of QPM-DWs. QPM has also been demonstrated for both modulation-instability as well as soliton-DW phase-matching using width-oscillating fibers \cite{kudlinski_parametric_2015, conforti_multiple_2016, droques_dynamics_2013} and for fiber-Bragg gratings \cite{westbrook_supercontinuum_2004, kim_improved_2006, zhao_observation_2009, yeom_tunable_2007, westbrook_perturbative_2006}. QPM has also been seen in Kerr frequency comb generation \cite{huang_quasi-phase-matched_2017}. On-chip waveguides provide a powerful new platform for QPM-DW generation, offering straightforward dispersion engineering, access to a range of modulation periods, scalable fabrication, and the ability to access well-defined higher-order modes.

In this first demonstration, the spectral brightness of the QPM-DWs was limited by several factors. First, most of the QPM-DWs were generated in the $\mathrm{TE_{20}}$ mode, which doesn't have optimal overlap with the $\mathrm{TE_{00}}$ mode. Indeed, in the case where a $-1$-order QPM-DW is generated in the $\mathrm{TE_{00}}$ mode (Fig. \ref{fig:suspended}a), the intensity of the light is higher. Second, for the cladding-modulated waveguides, we rely on a QPM structure with a high duty cycle, which effectively spreads the available QPM efficiency over many QPM orders, sacrificing efficiency in one particular order. Third, our waveguides only made modest changes to the effective mode area, and stronger QPM could be likely achieved with a deeper width modulation or stronger change in the cladding index. Finally, the QPM-DWs are often produced far from the pump wavelength, in a spectral region where the soliton is dim. In future designs, optimized strategies for QPM-DW generation could utilize somewhat longer modulation periods, allowing QPM to the $\mathrm{TE_{00}}$ mode, thereby maximizing mode-overlap and allowing the QPM-DWs to be located closer to the soliton central wavelength. Additionally, the waveguide modulation could be designed such that the efficiency of $\pm1$ order QPM is optimized. All of these parameters can be modeled using software that calculates the modes of the waveguide. This fact, combined with the massive scalability of lithographic processing, should allow for rapid progress in designing optimized photonic waveguides for SCG.

Currently, designers of waveguide-SCG sources work in a limited parameter space: selecting materials and selecting the dimensions of the waveguide cross section. QPM opens a new dimension in the design-space for photonic waveguides, one that is largely orthogonal to the other design dimensions. This orthogonality exists both in real-space, since the QPM-modulations exist in the light-propagation direction, but also in the waveguide-design-space, as it provides a simple vertical shift of the phase-matching conditions with no bending of the index curve (Fig.~\ref{fig:overview}c). Consequently, it allows the spectral location of DWs to be modified with minimal effect on the GVD at the pump wavelength, which enables the soliton propagation conditions to be controlled separately from the DW phase-matching conditions. For example, using QPM, DWs could be produced even for purely anomalous GVD, greatly relaxing the requirements for material dispersion and waveguide cross section. Importantly, since the GVD at the pump is known to affect the noise properties of the SCG process \cite{dudley_supercontinuum_2006}, the ability to manipulate the locations of the DWs separately from the GVD could enable SCG sources that are simultaneously broadband and low-noise. In addition, since similar phase-matching conditions apply to SCG with picosecond pulses or continuous-wave lasers \cite{dudley_supercontinuum_2006}, QPM of the SCG process is likely not restricted to the regime of femtosecond pulses.

In summary, here we demonstrated that quasi-phase-matching is a powerful tool for controlling the supercontinuum generation process in on-chip photonic waveguides. We experimentally verified that a periodic modulation of either the waveguide width or cladding can allow $\mathrm{Si_3N_4}$ waveguides to produce dispersive wave light at tunable spectral locations. By allowing dispersive waves to be quasi-phase-matched without significantly modifying the dispersion at the pump wavelength, this approach provides independent control over soliton compression and the spectral location of dispersive waves. Thus, quasi-phase-matching provides a new dimension in the design-space for on-chip waveguides and allows supercontinuum sources to be tailored for the specific needs of each application.

\begin{acknowledgments}
We thank Jordan Stone, Nate Newbury, Michael Lombardi, and Chris Oates for providing helpful feedback on this manuscript. We thank Alexandre Kudlinski for his insightful comments on a draft of this manuscript. This work is supported by AFOSR under award number FA9550-16-1-0016, DARPA (DODOS and ACES programs), NIST, and NRC. This work is a contribution of the U.S. government and is not subject to copyright in the U.S.A.
\end{acknowledgments}

\clearpage
\onecolumngrid
\begin{center}
	\textbf{\large Supplemental Materials:\\Quasi-phase-matched supercontinuum-generation in photonic waveguides}
\end{center}
\twocolumngrid
\setcounter{equation}{0}
\setcounter{figure}{0}
\setcounter{table}{0}
\setcounter{page}{1}
\makeatletter
\renewcommand{\theequation}{S\arabic{equation}}
\renewcommand{\thefigure}{S\arabic{figure}}

\maketitle

\section{Waveguide fabrication}

The width-modulated $\mathrm{Si_3N_4}$ waveguides were fabricated by Ligentec, using the ``Photonic Damascene'' process \cite{pfeiffer_photonic_2016}. They are fully $\mathrm{SiO_2}$-clad, and have a thickness of 750~nm, and a maximum width of 1500~nm. The end-sections are tapered to 150~nm at the input and exit facets in order to expand the mode and allow for improved coupling efficiency, which is typically --2~dB per facet. The overall length of the waveguides is 15~mm, and, over a 6-mm central region, the width is modulated sinusoidally from 1250 to 1500~nm. The modulation period ranges from 5.5 to \SI{6.5}{\micro\meter}.

The ``suspended'' cladding-modulated waveguides were fabricated at NIST Gaithersburg. First, 700~nm of Si$_3$N$_4$ was deposited onto thermally oxidized silicon wafers using low pressure chemical vapor deposition (LPCVD), and the waveguides were patterned with electron-beam lithography and reactive ion etching.  Plasma enhanced chemical vapor deposition (PECVD) was then used to selectively deposit a SiO$_2$ cladding on the input and output coupling regions. The sample was then coated with photoresist and patterned to define the suspended regions. After development, the sample was soaked for 30 minutes in a 6:1 solution of buffered oxide etch (BOE) to release the waveguides. The waveguides are 12~mm in total length, and the first and last 0.5~mm are clad with SiO$_2$ in order to provide a symmetric mode profile that improves coupling efficiency. The SiO$_2$ supports are on a pitch of \SI{200}{\micro\meter} and are in contact with the $\mathrm{Si_3N_4}$ waveguide for approximately \SI{20}{\micro\meter}.

\section{Experiment}
The light is coupled into each waveguide using an aspheric lens (numerical aperture of 0.6) designed for 1550~nm. The light is collected by butt-coupling an $\mathrm{InF_3}$ multimode fiber (numerical aperture of 0.26) at the exit facet of the chip. The waveguide output is then recorded using two optical spectrum analyzers (OSAs); a grating-based OSA (Ando 6315E) is used for the spectrum across the visible and near-infrared regions, while a Fourier-transform OSA (Thorlabs OSA205) extends the coverage to 5600~nm. The angle of the half-waveplate is controlled using a Thorlabs K10CR1 rotation stage.

\begin{figure}
	\includegraphics[width=\linewidth]{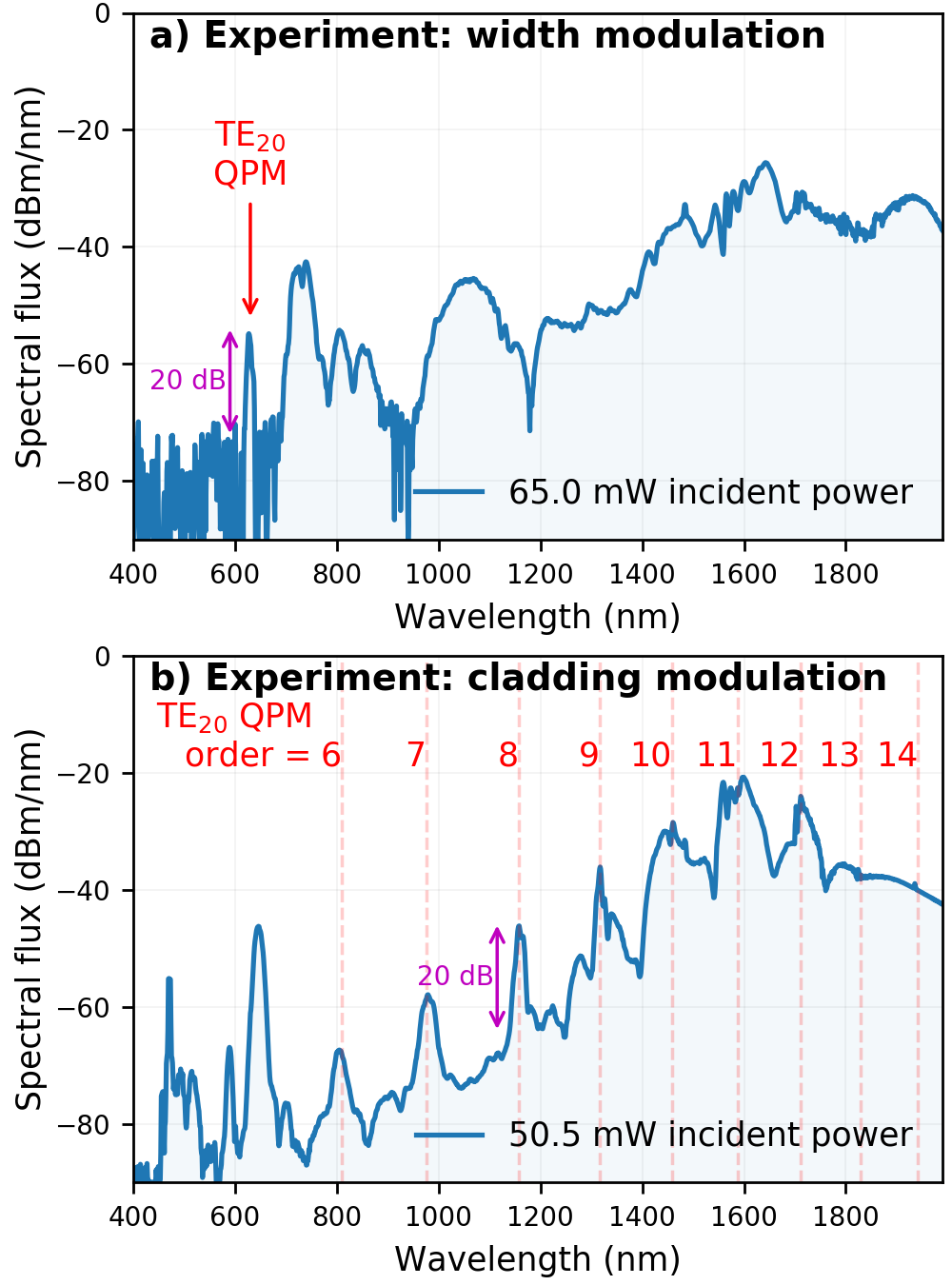}
	\caption{\label{fig:lineouts} Experimental supercontinuum spectra showing quasi-phase-matched dispersive-waves (QPM-DWs) due to waveguide modulation. a) Spectrum of the supercontinuum light produced from the width-modulated waveguide shown in Fig.~2a (6.2-\SI{}{\micro\meter} width-modulation period) with 65.0~mW incident power. Enhancement of the spectral flux to the $\mathrm{TE_{20}}$ mode can be seen as a sharp peak near 630~nm. b) Spectrum of the supercontinuum light produced by the cladding-modulated waveguide shown in Fig.~4a (width of 4000-nm) with 50.5~mW of incident power. Numerous QPM orders can be seen with peak heights of $>$20 dB in some cases. The dashed red lines show the theoretically predicted position of each $\mathrm{TE_{20}}$ QPM order. }
\end{figure}

For easy visualization of the spectral enhancement due to QPM, Fig.~\ref{fig:lineouts} shows representative spectra from both the width-modulated and cladding-modulated waveguides. These spectra are single rows of Fig.~2a and 4a in the main text and in each case, the QPM-enabled spectral peaks can have heights of 20~dB or more relative to the supercontinuum. 

\section{Waveguide modes}
We label the quasi-transverse-electric (TE, horizontal polarization) and quasi-transverse-magnetic (TM, vertical polarizaion) waveguides modes using subscripts that indicate the number of nodes in the $x$- and $y$-directions respectively (Fig.~3). For example, the $\mathrm{TE_{20}}$ mode (Fig.~3e) has two nodes in the $x$-direction and zero nodes in the $y$-direction. Experimentally, we observe quasi-phase-matched dispersive wave (QPM-DW) generation from the $\mathrm{TE_{00}}$ mode to the $\mathrm{TE_{20}}$, which seems somewhat counter-intuitive, since in a waveguide, all modes \textit{at the same wavelength} are orthogonal. However, it is possible for one mode at a certain wavelength to have a nonzero overlap with other modes at a different wavelength. For a simple rectangular waveguide with symmetric cladding (Fig.~3), overlap to the $\mathrm{TE_{00}}$ mode can only occur for symmetric modes, i.e., modes that have even numbers of $x$ and $y$ nodes (where zero is an even number). For example, if we consider the antisymmetric $\mathrm{TE_{10}}$ mode (Fig.~3c), we find that the mode overlap integral will be strictly zero, since any overlap on the right side will be precisely canceled by the antisymmetric left side. In contrast, the $\mathrm{TE_{20}}$ mode (Fig.~3e) can have nonzero overlap with the $\mathrm{TE_{00}}$ mode. Thus, to generate light into higher order modes using a lowest-order-mode pump, symmetric modes such as the $\mathrm{TE_{20}}$ mode offer the best overlap \cite{guo_second-harmonic_2016}.

\section{QPM mechanism}
The width- and cladding-modulated waveguides provide QPM for the DW phase-matching condition. However, the experimental results do not specify if the effect is due to the modulation of the intensity of the light (and therefore modulation of the effective nonlinearity of the waveguide, $\gamma$) or if it is due to the modulation of the waveguide group-velocity dispersion (GVD), which changes the effective phase-mismatch for soliton-DW phase-matching along the waveguide length. Depending on the specific situation, it is conceivable that either effect causes the experimentally observed QPM.

Thus, we employ numerical solutions to the Nonlinear Schr\"odinger equation (NLSE) \cite{heidt_efficient_2009, hult_fourth-order_2007,ycas_pynlo_2016} to investigate which effect plays the most important role for our waveguides. We run NLSE simulations in three cases: 
\begin{enumerate}
	\item Including the modulation of both the GVD and $\gamma$, 
	\item Including the modulation of $\gamma$ only, 
	\item Including the modulation of the GVD only. 
\end{enumerate}
Our simulations only consider the fundamental mode of the waveguide and do not take into account effects due to higher order modes, which prevents them from modeling the QPM-DWs in the $\mathrm{TE_{20}}$ mode. 

\begin{figure*}
	\includegraphics[width=\linewidth]{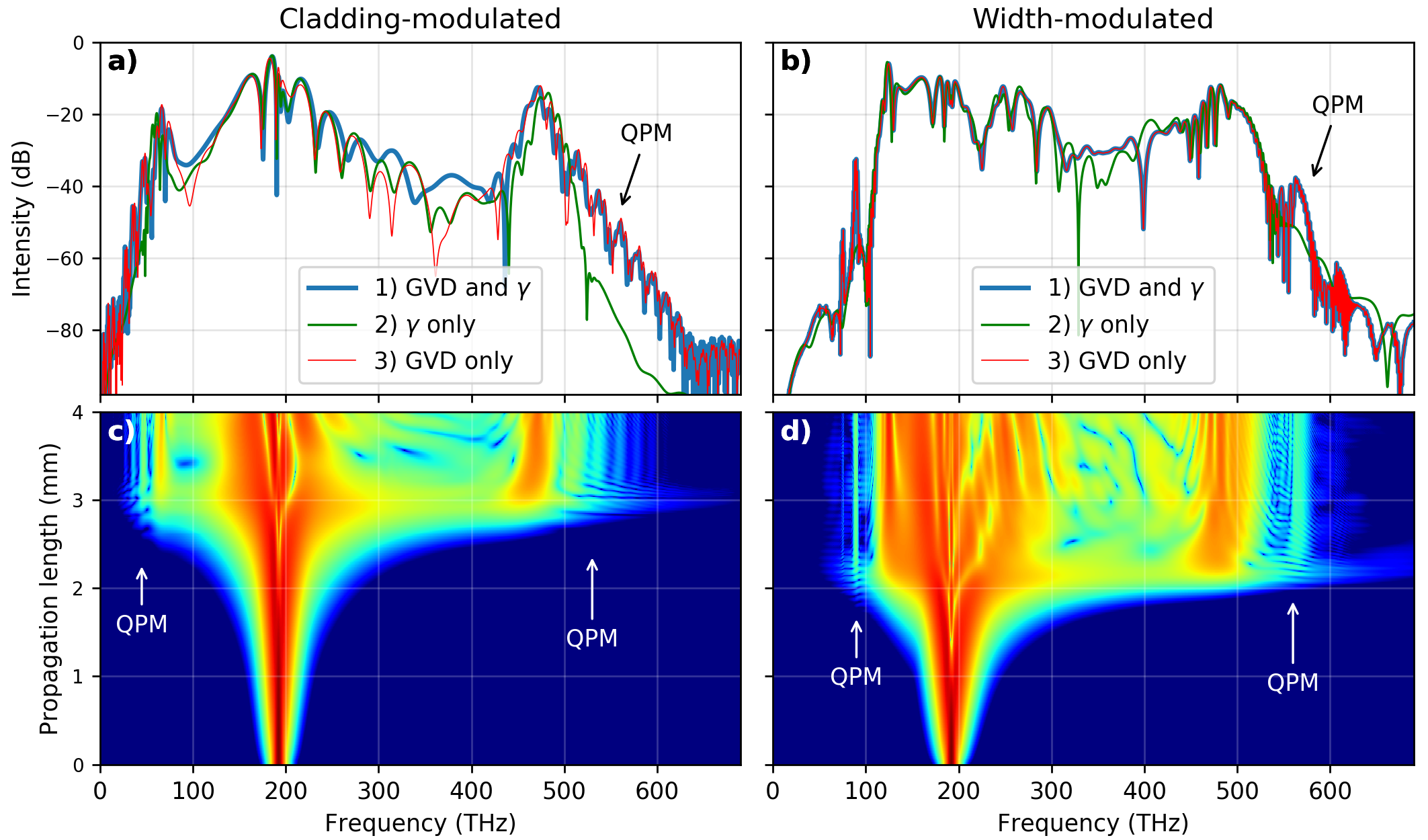}
	\caption{\label{fig:NLSE} Numerical simulations using the Nonlinear Schr\"odinger Equation (NLSE) for 80-fs, 300-pJ pulses propagating through cladding-modulated (left) and width-modulated (right) waveguides. a,b)~QPM effects can be seen in the case where modulations to the group velocity dispersion (GVD) alone are included and in the situation where modulation of the GVD and the effective nonlinearity ``GVD and $\gamma$'' are included. However, only including modulations to $\gamma$ is not sufficient to achieve QPM. c,d) The spectrum as a function of propagation length along the waveguide. In the region of soliton fission, the intensity of the QPM-DWs oscillates as a function of propagation length. \textbf{Simulation parameters:} The cladding-modulated waveguides are modeled using values from the experiment: a thickness of 700~nm, a width of 4000~nm, with periodic $\mathrm{SiO_2}$ bottom-cladding with a length of \SI{20}{\micro\meter} on a \SI{200}{\micro\meter} pitch. The width-modulated waveguides are modeled using height (650 nm) and width (sinusoidal variation between 1250 and 1500 nm) values from experiment, but with a modulation period of \SI{80}{\micro\meter} in order to show QPM-DW generation to the $\mathrm{TE_{00}}$ mode.)}
\end{figure*}

Fortunately, in the case of the cladding-modulated waveguides, QPM-DWs to the fundamental ($\mathrm{TE_{00}}$) mode are experimentally observed on the short-wavelength side of the main DW (Fig.~4a). These features are reproduced by the NLSE (Fig.~\ref{fig:NLSE}a,c) \emph{only} when the modulation of the GVD is taken into account. When the simulations include the modulation of $\gamma$ alone, no obvious QPM effects are seen, suggesting that for the cladding-modulated waveguides, phase-matching is enabled by GVD modulation.

For the width modulated waveguides, the only experimentally observed QPM-DWs are those in the $\mathrm{TE_{20}}$ mode, and thus we do not attempt to precisely model the experiment, since the simulations do not take into account higher order modes. Instead, we model a waveguide with a similar sinusoidal width modulation (1250 to 1500~nm), but with a modulation period of \SI{80}{\micro\meter}, which enables a QPM-DW in the $\mathrm{TE_{00}}$ mode. Again, the NLSE indicates that the modulation of the GVD is the dominant effect (Fig.~\ref{fig:NLSE}b,d), in agreement with previous studies of QPM-DW generation \cite{kudlinski_parametric_2015, conforti_multiple_2016, droques_dynamics_2013}. We expect that the GVD will also provide QPM for the QPM-DWs in the $\mathrm{TE_{20}}$ mode.

This method of QPM via GVD modulation can be understood by considering that the soliton and the DW are phase-mismatched, and that the GVD determines the degree of mismatch. As a result, the DW light experiences regions of constructive and destructive interference along the length of the waveguide. By modulating the GVD, the regions of constructive interference can be made slightly longer than the regions of destructive interference. The net result is that, while the intensity of the DW still oscillates to some degree, there is a constructive build-up of light intensity \cite{droques_dynamics_2013}. The oscillation and build-up of the QPM-DW light as a function of propagation length can be seen in Figs.~\ref{fig:NLSE}c and \ref{fig:NLSE}d in a region near the point of soliton fission (approximately 3~mm in Fig.~\ref{fig:NLSE}b and 2~mm in Fig.~\ref{fig:NLSE}d).

\section{Disclaimer}
Certain commercial equipment, instruments, or materials are identified here in order to specify the experimental procedure adequately. Such identification is not intended to imply recommendation or endorsement by the National Institute of Standards and Technology, nor is it intended to imply that the materials or equipment identified are necessarily the best available for the purpose. This work is a contribution of the United States government and is not subject to copyright in the United States of America.

\bibliography{Zotero}

\end{document}